\documentstyle[emulateapj]{article}
\lefthead{I. Schmoldt, P. Saha}
\righthead{Variational Dynamics and the Local Group}
\begin{document}

\title{On Variational Dynamics in Redshift Space}
\author{Inga M. Schmoldt and Prasenjit Saha}
\affil{Department of Physics (Astrophysics), Oxford University, Keble Road, 
       Oxford OX1 3RH, UK}

\begin{abstract}
\cite{peeb1} showed that in the gravitational instability picture
galaxy orbits can be traced back in time from a knowledge of their
current positions, via a variational principle. We modify this
variational principle so that galaxy redshifts can be input instead of
distances, thereby recovering the distances.
As a test problem, we apply the new method to a Local Group model. We
infer $M = 4$ to $8 \times 10^{12} M_{\sun} $ depending on cosmology,
implying that the dynamics of the outlying Local Group dwarves are
consistent with the timing argument.
Some algorithmic issues need to be addressed before the method can be
applied to recover nonlinear evolution from large redshift surveys,
but there are no more difficulties in principle.

\end{abstract}


\section{Introduction}

Phase space is six--dimensional and therefore six numbers for each
particle will specify the
dynamics completely. The standard approach is to define initial
conditions as six
numbers (positions and velocities in three dimensions) and integrate
those forward in time. This is the usual $N$-body approach to the problem.

Alternatively, it is not necessary to define those six numbers at only
one time. It is equally well possible to split them, such that three
numbers will be known at an initial time, and three at a final time,
where the former are derived from physical arguments about the inital
state, and the latter are provided by data on the current state of the
system. This is the boundary value approach.

The usual boundary value for structure formation by hierarchical
clustering at initial times is the gravitational
instability requirement that initial peculiar velocities must vanish:
therefore, it is possible to express the orbits as a sum of growing
modes. This is what perturbation theory is designed to do, and
linear perturbation theory and its extension, the Zel'dovich
approximation are in wide use.
However, as structure formation is non-linear on small scales, a
non-linear formalism would be more useful. Non-linear perturbation
theory exists (e.g. \cite{nus}, \cite{gram93a}, \cite{gram93b},
\cite{buch}) but is very complicated and has convergence problems,
i.e. the dynamics at early times can be fitted to a high accuracy only
at the expense of a good fit at later times.

A different method (cf \cite{peeb1}) of addressing the boundary value
problem is to use the variational principle. The basic method is to
start with a 
parameterisation of the orbits which satisfies the boundary
conditions and then adjust the parameters until a stationary point of the
action is found. (A variant, suggested by \cite{gia93} and implemented
by \cite{susp}, parameterises the density and velocity fields rather
than the orbits.)
This may seem like perturbation theory because it is
a method which attempts to get better and better approximations of
galaxy orbits but there are important differences. The main one is
that perturbation theory attempts to fit early times even at the
expense of accuracy at later times, whereas the variational principle
spreads out the errors more uniformly over all times. The variational
principle is also algorithmically more straightforward to do to higher
orders.

The main disadvantage of Peebles' original method is that it requires
the input of distances to recover redshifts. Redshifts, however, are
easy to measure, whereas distances are not. We therefore change to
recovering distances from redshifts.
The recent
nearly all-sky redshift surveys QDOT and PSC$z$ provide a strong
motivation for variational methods in redshift space.  
\cite{shay93} and \cite{shay95} have pointed out that this could be
achieved by treating the 
distance boundary condition as a parameter and fitting that until the
recovered redshifts agree with the measured ones. 
Another approach is to modify the variational principle until the
boundary conditions are of the desired form. We take such an approach
and so does \cite{whit98} but the details in his treatment differ
from ours. The attractive feature of this approach is that it only
requires small modifications of Peebles' elegant original method.

In this paper we develop a redshift space variational method, apply
it to a small system --- the Local Group ---, 
and point out the algorithmic problems that need to be solved to extend
to larger systems.

Various aspects of the Local Group dynamics have been studied by \cite{peeb1},
\cite{peeb2}, \cite{peeb4}, and \cite{dunn}. The new feature of our
analysis is that we attempt to constrain the
masses of the Milky Way, M31, and an underlying distribution of
unclustered matter by using a likelihood approach.

Two points need to be made about these variational methods in
general: Firstly, the true solutions
of the variational action need not be a minimum even though the
colloquial use of `least action' has stuck (cf \cite{peeb2}). 
Secondly, the equations
solved are the same as for $N$-body integration, only the
approximations used to solve them are different. In neither of the
two approaches do particles have to be galaxies -- they may well be
samplers of some underlying distribution function and as 
\cite{branch} and \cite{dunn2} have pointed out, applying the
variational method using galaxies only, neglecting biasing and
unclustered matter, leads to wrong results.

\section{Formulation}
\label{sec:maths}

Orbits of galaxies follow equations of motion of the form
\begin{equation}
\frac{d}{dt}(a^2 \dot{\bf{x}}) = - \frac{\partial \Phi}{\partial {\bf x}},
\label{eq:motion}
\end{equation}
where $\bf{x}$ are some comoving coordinates, and $a(t)$ is the
scale factor of the universe. These equations are
derivable from a stationary action, $\delta S = 0$, subject to some boundary
conditions. In this work, we will distinguish between two different
actions and different sets of boundary
conditions leading to the same equations of motion. 

\subsection{Two Kinds of Boundary Condition}

For simplicity,
let $x$ be one-dimensional in this subsection. Then the two cases are:
\begin{itemize}
\item the `real space' case, where the boundary conditions are 
$a^2 \dot{x} = 0 $ at $t = 0$ (i.e., initial peculiar velocities zero)
and $\delta x = 0$ at $ t = 1$ (i.e., present positions
fixed) with the action given by
\begin{equation}
S = \int_{0}^{1} L dt {\hskip 5 pt} \mbox{, where} {\hskip 5 pt}
L = \hbox{$1 \over 2$} a^2 \dot{x}^2 - \Phi(x)
\end{equation}
This is the action and the boundary conditions first proposed by \cite{peeb1}.
\item the `redshift space' case, where the initial boundary condition
remains the same and the final one becomes $ \dot{a} \delta x + a
\delta \dot{x} = 0 $ at $ t = 1$ (i.e. present redshifts are fixed). 
The new boundary conditions will still
lead to the same equations of motion if we change the action to
\begin{equation}
S = \int_{0}^{1} \left( L - \dot{G} \right) dt \mbox{  , where  }
G = a^2 x \dot{x} + \hbox{$1 \over 2$} a \dot{a} x^2
\end{equation}
and $L$ remains the same. \cite{whit98} describes more general
transformations of this type.
\end{itemize}

The standard method for finding an orbit that will keep the action
stationary and therefore is a solution to the equations of motion
consists of formulating a parametric expression for the coordinates
$x$ which satisfies the boundary conditions and adjusting the
parameters to make the action stationary. For the real space case, we choose
the expression
\begin{equation}
x = \sum_{n=1}^N C_{n} f_n(t),
\label{eq:x12}
\end{equation}
where $C_{n}$ are a series of $N$ coefficients and $f_n(t)$ is a
temporal basis function.
For the redshift space case,
\begin{equation}
x = \frac{cz}{\dot{a}(1)} + \sum_{n=1}^N C_{n} g_n(t)
\end{equation}
To ensure that the boundary conditions are fulfilled, the basis
functions have to satisfy
\begin{eqnarray}
f_n(1) &=& 0 \nonumber \\
g_1(t) &=& f_1(t) - \frac{\dot{f}_1(1)}{\dot{a}(1)} 
\mbox{, }{\hskip 12pt} g_{n > 1}(t) = f_n(t)
\end{eqnarray}

A good choice for $f_n$ is $(1-D(t))^n$, where $D(t)$ is the growth
factor from linear theory; this was proposed by \cite{gia93}. In this
case, for $N=1$ the variational method reduces to linear theory. But
all that is essential is that linear theory should be followed as $t
\rightarrow 0$. So, other choices such as $f_n = (1- t^{2/3})^n$ are
also possible.

\subsection{Equations for Stationary Action}
The real three-dimensional problem is a combination of the two types
of boundary conditions. Because of the way in which we choose coordinate
systems, the real space treatment can be
used for the first two dimensions ($x$, $y$) of each orbit and only the third
dimension ($z$) will have to be treated in redshift space. We will
refer to axes by their dimension rather than by a $x$, $y$, or $z$ in
order to prevent confusion between the third dimension and the symbol
used for redshifts.

One coordinate system is assigned to each object and all coordinate systems
share a common origin. This origin is the point from which a comoving
observer would measure the objects' redshifts. It is important to note
that, although the objects move, their coordinate frames do not.
By pointing the $3$-axis of each frame towards the current galactic
coordinates $l$,$b$ of the object associated with it, we
ensure that the radial velocity is along one axis only. Therefore, the
redshift space treatment will have to be applied only to that axis.
With this orientation, the $3$-coordinate of each object is 
$cz_i$ ($cz_i = \dot{a} x_{i_3} + a \dot{x}_{i_3}$),
where $z_i$ is the
`comoving' redshift, i.e. the redshift measured by an observer who is at 
rest with respect to the microwave background. The
$1$,$2$-coordinates at $t = 1$ are 
$ x_{i_1} = 0 \mbox{, } x_{i_2} = 0$, so the current positions are
completely specified and real space treatment can be applied.

In these coordinates, the parametric expansion of the position $\bf{x}$
of an object $i$ becomes
\begin{equation}
x_{i_{(1,2)}} = \sum_{n=1}^N C_{i_{(1,2)},n} f_n(t)
\end{equation}
\begin{equation}
x_{i_{(3)}} = \frac{cz}{\dot{a}(1)} + \sum_{n=1}^N C_{i_{(3)},n} g_n(t)
\end{equation}
and the full action is $\int_0^1 (L-\dot G)$, where $L$ and $G$ are
\begin{equation}
L = \hbox{$1 \over 2$} \sum_i m_i a \dot{{\bf x}}_i^2 - 
\hbox{$1 \over 2$} \ddot{a} a
\sum_i m_i {\bf x}_i^2 + \frac{1}{a} \sum_{i,j} \frac{m_i m_j}{| {\bf
x}_i - {\bf x}_j |^{1/2}} - \Phi_{\rm tidal}
\label{eq:L}
\end{equation}
and
\begin{equation}
G = \sum_i m_i \left( a^2 x_{i_{(3)}} \dot{x}_{i_{(3)}} + \hbox{$1 \over 2$}
a \dot{a} x_{i_{(3)}}^2 \right) .
\label{eq:G}
\end{equation}

The first term in equation~\ref{eq:L} describes the total kinetic
energy, the second term is an acceleration caused by the fact that the
coordinate system is expanding at a rate $a(t)$, and the third term
describes the gravitational interaction between members of the group
of objects.  
$\Phi_{\rm tidal}$ represents the tidal potential caused by the
influence of objects external to the region considered.
Note that introducing
a homogeneous mass distribution into the system is very easy
since it will only rescale the second term.

Inserting the  equations ~\ref{eq:L} and ~\ref{eq:G} into the action $S$
and taking the gradient with respect to the $C_i$ leads to
\begin{eqnarray}
\left[ \int_{0}^{1} a^2 \dot{f}_m \dot{f}_n dt \right] C_{i_{(1,2)},n}
= \nonumber \\
\int_{0}^{1} \frac{f_n}{a} \sum m_i m_j \nabla_{i_{(1,2)}} |{\bf
x}_{i} - {\bf x}_{j} |^{-1} - \frac{\partial\Phi_{\rm tidal}}
{\partial C_{i_{(1,2)}}}
\label{reals}
\end{eqnarray}
and
\begin{eqnarray}
&&\frac{\dot{f}_1(1)^2}{\dot{a}(1)} C_{i_{(3)},1} + \left[
\int_{0}^{1} a^2 \dot{g}_m \dot{g}_n dt \right] C_{i_{(3)},n}
\nonumber \\
&&+\int_{0}^{1} \dot{a} a g_n dt \frac{cz_i}{\dot{a}(1)} = \nonumber \\
&&\int \frac{g_n}{a} \sum m_i m_j \nabla_{i_{(3)}} |{\bf x}_{i} - {\bf
x}_{j} |^{-1} - \frac{\partial\Phi_{\rm tidal}}
{\partial C_{i_{(3)}}}
\label{reds}
\end{eqnarray}

To find the orbits which keep the action stationary, we now have to
solve equations~\ref{reals} and ~\ref{reds} for the spatial
coefficients $C_{i,n}$. The standard method
is to assume some initial coefficients, calculate the right hand
sides, then solve the left hand
sides for the coefficients and use the results to recalculate the right
hand sides. Note that the left hand side integrals in equations
\ref{reals} and \ref{reds} have to be done only once. 

We have experimented with various procedures to help
convergence, e.g. adding a stabilising term 
$ \propto \left[ \int \ddot{a} a f_m f_n dt \right] C_{i,m}$ on 
both sides of the equation or scaling down some terms
for early iterations and only slowly increasing them to their full
value (cf \cite{susp}, \cite{gia93}). 

\subsection{Units}
It is convenient to define model units for time, mass, and length (tick,
marble, and lap) which we keep scaleable with the age of the universe
$T_0$, since $T_0$ is not known. The units are given by
\begin{equation}
1 \mbox{ tick} = T_0 = 10 {\hskip 3pt} \kappa {\hskip 3pt} {\rm Ga}
\end{equation}
and $\kappa$ is the age of the universe in units of 10
Ga. We do not want the velocities to be scaleable, so we require $100
\mbox{ km/s} = 1 \mbox{ lap/tick}$ which leads to
\begin{equation}
1 \mbox{ lap}  = 1.023 {\hskip 3pt} \kappa {\hskip 3pt} {\rm Mpc}
\end{equation}
The assumption of a gravitational constant equal to unity as in the
equations above leads to 
\begin{equation}
1 \mbox{ marble} = 2.38 \cdot 10^{12}{\hskip 3pt} \kappa {\hskip 3pt}
M_{\sun}
\end{equation}
We therefore have a one-parameter family of models for different ages
of the universe.

\section{Application to the Local Group}
\label{sec:data}
In this section we will apply the above formalism to a
small group of galaxies, namely the Milky Way--M31 system and outlying
Local Group dwarves, with external forces approximated by dipole and
quadrupole tidal forces growing according to linear theory. 
Since distances to
objects within this group are fairly well known, we can constrain the mass
distribution by comparing the distances predicted by our method to the
observed distances for a range of different mass distributions. We also test
the effects of assuming different values for $\Omega_0$.

\subsection{Formulation for the Local Group}
The formalism is outlined in section \ref{sec:maths}, but
for the Local Group there are two special considerations:

Firstly, the Milky Way has a special role.  As explained above,
each galaxy has its own coordinate system, and we take the current
position of the Milky Way as the common origin.  This origin is at
rest with respect to the CMB and for $t\neq1$ the Milky Way moves away
from it.  As the Milky Way's current position is fixed the real space
treatment can be applied in all three dimensions.

Secondly, cosmology only influences the Local Group
via tidal forces. Therefore, we can keep things simple by introducing
the following complication:
The integrals on the left hand sides of 
equations ~\ref{reals} and ~\ref{reds} are particularly simple to
solve for $a(t) = t^{2/3}$, and we exploit this fact 
by setting $a(t) = t^{2/3}$ regardless of cosmologies.  This $a(t)$ is
only the scale factor of a convenient frame that we have
chosen for the Local Group and which need not necessarily bear any
relation to the real physical scale factor
of the universe $a_{\rm phys}(t)$ except at the boundaries
\begin{eqnarray}
&&a(t \rightarrow 0) = t^{2/3} \propto a_{\rm phys}(t \rightarrow 0)
\nonumber \\ 
&&a(t = 1)  = 1  = a_{\rm phys} (t = 1)
\end{eqnarray}
For interactions between members of the group, the value of $a(t)$
does not matter. Only when calculating the influence of the tidal
forces do we have to consider the relation between the two sets of
comoving coordinates ${\bf x}_{\rm phys} = \frac{a}{a_{\rm phys}}
{\bf x}$. The tidal potential then is
\begin{equation}
\Phi_{\rm tidal} = \frac{D(t)}{a_{\rm phys}} \left[ {\bf d} \cdot {\bf
x}_{\rm phys} + \hbox{$1 \over 2$} {\bf x}_{\rm phys} \cdot {\bf Q} \cdot {\bf
x}_{\rm phys} \right]
\label{tides}
\end{equation}
where $D(t)$ is the growth rate of the density
fluctuations, ${\bf d}$ is the dipole and ${\bf Q}$ is the quadrupole 
strength.

The quadrupole potential is taken from
\cite{rl}. 
The dipole
is inferred indirectly from the Milky Way's known velocity with
respect to the microwave background, in the following way. As the
equations (11) and (12) are being
solved for the orbits of all galaxies involved, we
constantly monitor the development of the predicted velocity of the
Milky Way. This velocity depends on the strength of the dipole. During
the iterations, we fit that dipole strength such that the predicted
velocity of the Milky Way agrees with the one measured by
\cite{kogut}. 

\subsection{Finding Solutions in the Local Group}
We model the Local Group as an ensemble of 13 galaxies with the mass
of the system distributed between the Milky Way, M31,and some local
unclustered matter 
$\rho_{\rm sm}$.  It is convenient to express this $\rho_{\rm sm}$ in
units of the 
critical density, but it is important to note that it is only local
and does not change the cosmology.  The cosmological models have
$\Lambda=0$ and various $\Omega_0$.  The ratio of masses
of Milky Way and M31 is fixed at 2:3
and all dwarf galaxies are treated as test particles.

We select the dwarf galaxies by choosing all objects from the list of
Local Group members (\cite{hodge}) which are more than 500 kpc away from both
M31 and the Milky Way. This high distance is a necessary restriction
because the orbits of nearby galaxies
tend to be too dominated by the internal dynamics of the Milky
Way--M31 system and are therefore difficult to reconstruct. Nearby
galaxies might also be part of the system's halo.
The dwarf galaxies chosen, on the other hand, are generally not too
sensitive to the mass ratio between M31 and the Milky Way because of
their relatively great distance to that system.
Distances (from \cite{hodge}) and redshifts (from the Nasa/IPAC
Extragalactic Database) of all galaxies are listed in table 1.
Note
that these heliocentric redshifts are converted to CMB--centric
redshifts to get the appropriate boundary conditions.
{\vskip 5pt}
\placetable{tbl-1}
{\small
\begin{tabular}{lrrrrrr}
\tableline
\tableline
\noalign{\vskip3pt}
 Galaxy & $l$ & $b$& $cz$& $d_{\rm obs}$
& $d_{\rm model}$ \\
\noalign{\vskip2pt}
        & deg & deg & $\frac{{\rm km}}{{\rm s}}$  & kpc & (kpc/$\kappa$)\\
\noalign{\vskip3pt}
\tableline
\noalign{\vskip3pt}
M31        & 121.2 & -21.6  & -300  &  725 &   787  \\
IC1613     & 129.8 & -60.6  & -234  &  765 &   498  \\
WLM        &  75.9 & -73.6  & -116  &  940 &  1224 \\
Sextans A  & 246.2 &  39.9  &  324  & 1300 &  1878 \\ 
N3109      & 262.1 &  23.1  &  403  & 1260 &  1977 \\
IC10       & 119.0 &  -3.3  & -344  & 1250 &  1227 \\
Pegasus    &  94.8 & -43.5  & -183  & 1800 &  1911 \\
Sextans B  & 233.2 &  43.8  &  301  & 1300 &  2023 \\
SagDIG     &  21.1 & -16.3  &  -77  & 1150 &  1266 \\
LGS 3      & 126.8 & -40.9  & -277  &  760 &  1220 \\
EGB0427+63 & 144.7 &  10.5  &  -99  &  800 &  2069 \\
N6822      &  25.3 & -18.4  &  -57  &  540 &  1554 \\
\noalign{\vskip3pt}
\tableline
\end{tabular}
{\vskip3pt}
Table 1: Local Group Members; predicted distances are from the
solution pictured in figure 1}
{\vskip 5pt}

In order to increase our chances of finding the right solutions among
the many possible ones, we start the code with an initial guess by choosing
the $C_{i,n} = 0$ except for 
\begin{equation}
C_{i_{3},1} = d_{i_{\rm obs}} -
\frac{cz_i}{\dot{a}(1)}.
\end{equation}

This produces an initial set of ${\bf x}_i(t)$ which fit the redshifts and 
measured distances (if $1 \mbox{ lap} = 1 \mbox{ Mpc}$) but
which are not solutions of the equations of motion. 
We therefore add an extra term to 
equations ~\ref{reals} and ~\ref{reds} to change them into
something that the ${\bf x}_i$ are solutions of. This term is then
gradually removed from the equations during iteration. 

Our current
method always produces convergence of the coefficients, but does not
do so in a very 
efficient way. In most cases, we need several hundred iterations to
achieve convergence, where the main problem seems to be that some of the
converging $C_{i,n}$ are caught in a cycle of two values.
We have tried to remedy this problem in a very crude way and
were successful but only at a severe cost in iterations; this is one
of the problems that will 
have to be solved in a more general way, before the code can be
applied to large datasets.

We check our reconstructed ${\bf x}_i$ by taking their values at $t =
0.01$ as the initial conditions of an $N$-body integration.

Figure 1 shows the variational orbits and $N$-body orbits for a
system of a total mass of $3.10 \mbox{ marbles}$. 
For the mass distribution of table 1, the dipole acceleration
inferred is $0.85 \mbox{ } {\rm laps}/{\rm tick}^2$ in the direction
$l = 274^{\circ}$ and $b = 29^{\circ}$ (the direction is
indistinguishable from \cite{yahil}). 

\placefigure{fig-1}
\input epsf
\def\figureps[#1,#2]#3.{\bgroup\vbox{\epsfxsize=#2
    \hbox to \hsize{\hfil\epsfbox{#1}\hfil}}\vskip12pt
    \small\noindent Figure#3. \def\par{\endgraf\egroup\vskip12pt}}

\figureps[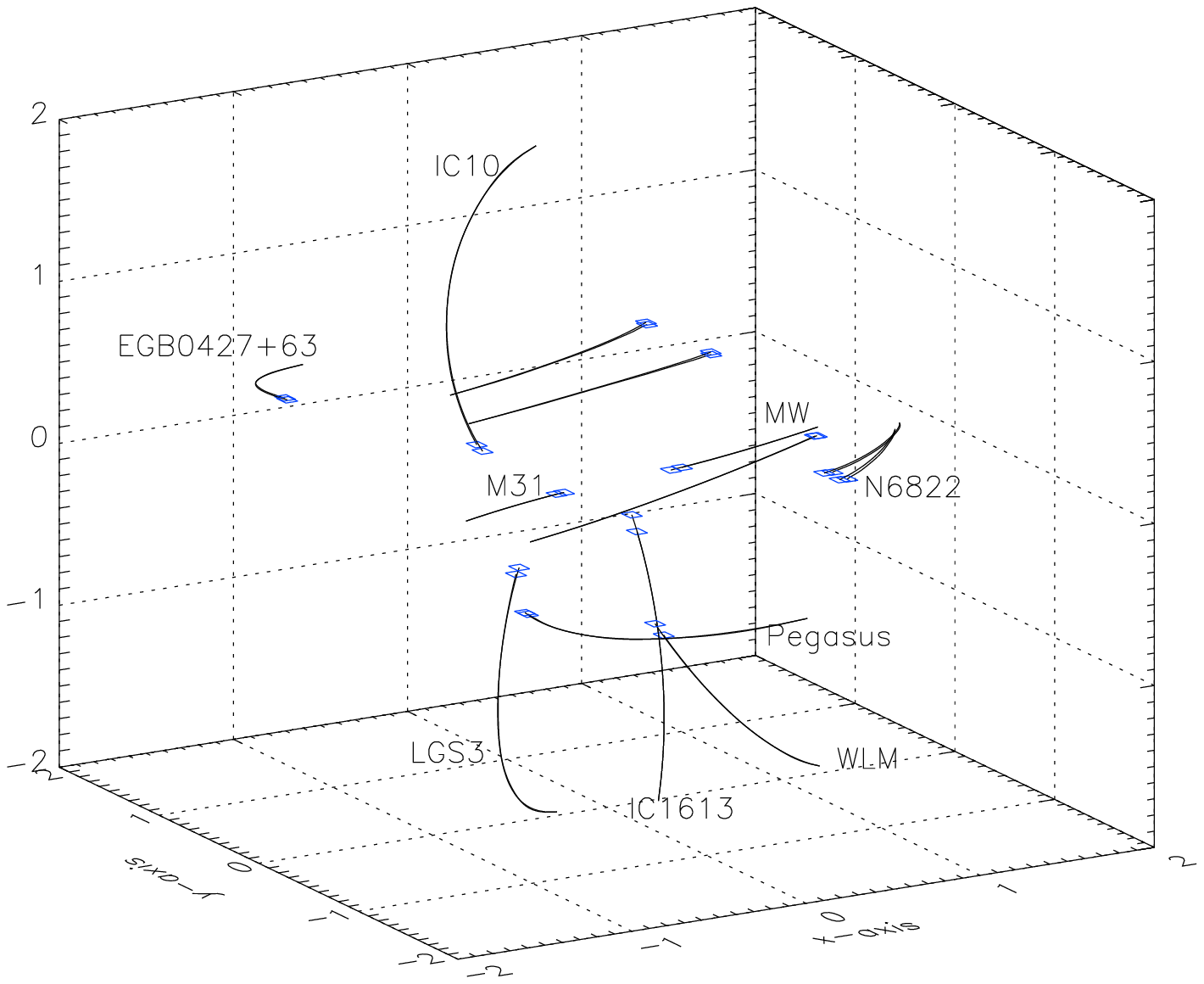,.9\hsize] 1. Variational and $N$-body
orbits for $\Omega = 0.99$, $m_{\rm tot} = 3.10 \mbox{ marbles}$ (no
unclustered matter); squares indicate positions at $t = 1$;
the scale is in comoving laps and the coordinate directions are
$(\cos b \cos l,\cos b \sin l,\sin b)$. We have labelled as many of
the galaxies as was possible without overcrowding the figure.

The last column of table 1 lists the predicted distances for this
particular model. They are generally similar to the measured
distances, which 
is not a trivial result, since as far as the mathematical
formalism is concerned, the predicted distances are not even
constrained to be positive. In fact, when trying to reconstruct the
orbits of galaxies which are too close to M31 or the Milky Way, the
code does produce negative distances.
A typical example
of this is Leo I (cf \cite{zar}): the code is given a positive
redshift but only finds approaching solutions. The only way to
reconcile these two requirements is 
to put the dwarf galaxy on the other side of the
Milky Way - which is what produces the negative distance. 
The reason for this is that nearby galaxies are too dominated by halo
dynamics and we have not attempted to model the halo in any way. Hence
our decision to exclude all the nearer dwarf galaxies.

\subsection{Analysis of Local Group Solutions}
To quantitatively analyse our results, we assign a likelihood to each
set of solutions and therefore to each mass distribution. 
The following is very crude, but we decided
to use it because it is well defined and uses plausible reasoning. The
likelihood is calculated in the
standard Bayesian fashion by assessing the probability of a set of
solutions given the set of measured distances. We have
to allow for the uncertainties associated with the measured distances,
and the fact that some of our solutions might simply be wrong (we call
this the `outlier probability'). As we have no particular reason to
believe that the uncertainties in $d_{\rm obs}$ are Gaussian, we use (for
all galaxies except M31) the
hatbox function
\[ \mbox{hb}(x,a,b) = \left\{ \begin{array}{ll}
	(b-a)^{-1} & \mbox{if $a \le x \le b$} \\
	0          & \mbox{otherwise}
	\end{array}
	\right. \]
If the outlier probability is $\alpha$ and the uncertainty in $d_{\rm obs}
$ is $\beta$ then 
\begin{eqnarray}
&&\mbox{prob}(d_{\rm obs}^i \mid d_{\rm model}^i) = \alpha
\mbox{hb} \left( d_{\rm obs},{\rm min}(d_{\rm obs}),{\rm max}(d_{\rm
obs}) \right)\nonumber \\ 
&&+ (1-\alpha) \mbox{hb}\left( d_{\rm obs}^i,
(1-\beta)d_{\rm obs}^i, (1+\beta)d_{\rm obs}^i \right)
\end{eqnarray}
For M31 we take $\mbox{prob}(d_{\rm obs}^i \mid d_{\rm model}^i)$
to be Gaussian where the observed distance has an associated
uncertainty of 5 \%. 

Combining all galaxies, we have,
\begin{equation}
\mbox{prob(data $\mid$ model,$\alpha$,$\beta$)} = \prod_i \mbox{prob}
(d_{\rm obs}^i \mid d_{\rm model}^i,\alpha,\beta)
\end{equation}

Since we do not know $\alpha$ or $\beta$, we marginalise by integrating
over a plausible range of both with a suitable prior, i.e. we
integrate over $0.1 \le \alpha \le 0.3$ with a flat prior and $0.1 \le
\beta \le 0.3$ with a $1/\beta$ prior. 

Figure 2 shows likelihood contours for two different values of
$\Omega_0$.
{\vskip 5pt}
\placefigure{fig-2}
{\vskip 5pt}
\input epsf
\def\figureps[#1,#2]#3.{\bgroup\vbox{\epsfxsize=#2
    \hbox to \hsize{\hfil\epsfbox{#1}\hfil}}\vskip12pt
    \small\noindent Figure#3. \def\par{\endgraf\egroup\vskip12pt}}

\figureps[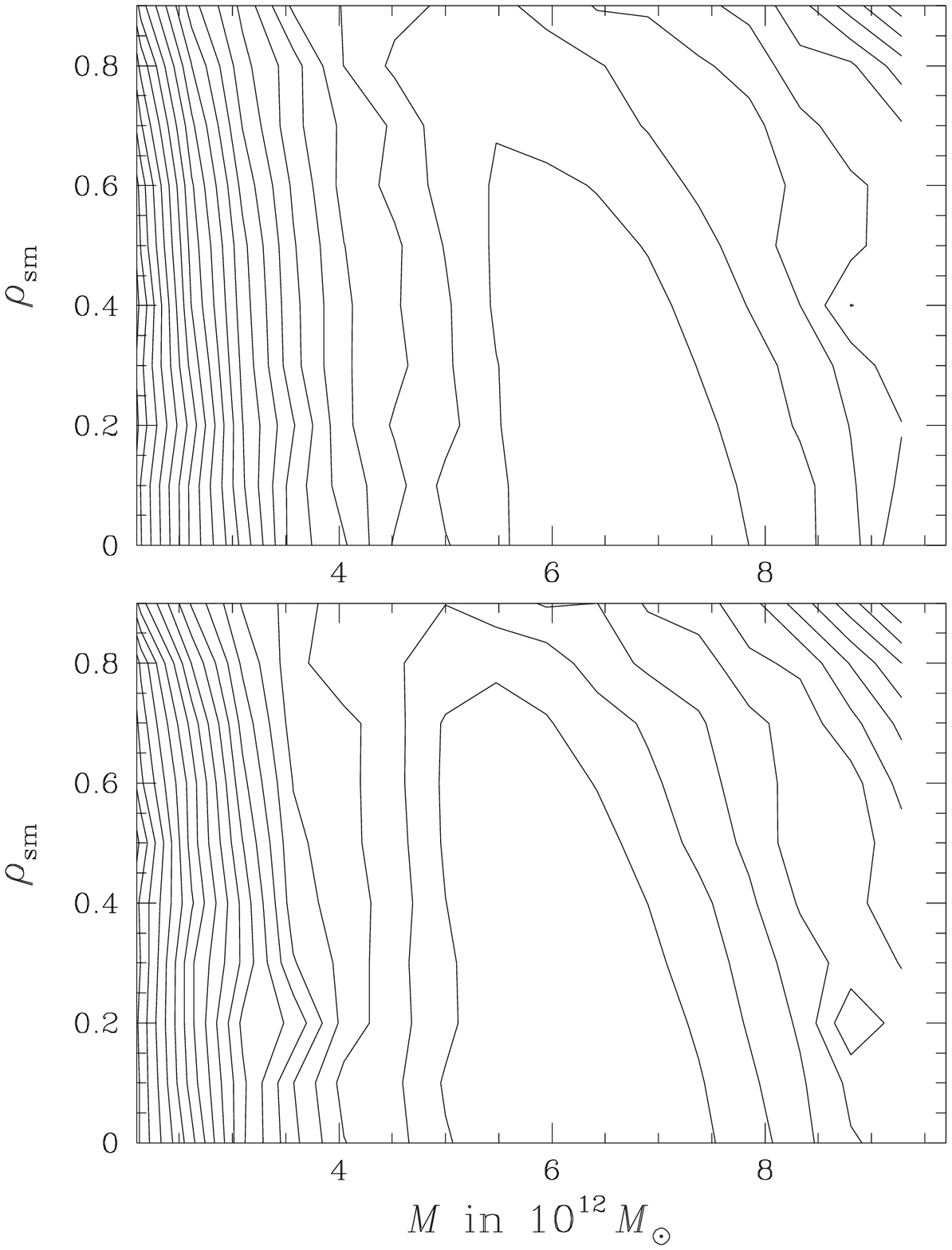,.9\hsize] 2. Likelihood contours for
variational solutions associated with different mass
distributions: above $\Omega = 0.4$, below $\Omega = 0.99$ 
In both cases, $\kappa = 1$. Contour levels differ by a factor of 10.

Masses are particularly high for low $\Omega_{0}$ because the
quadrupole force, which pulls the Milky Way and M31 apart is higher in
those cosmologies and so we require more mass to keep the two
galaxies at the same distance from each other.

The likelihood contours are dominated by M31:
leaving one or more of the dwarf galaxies out of the likelihood
analysis does not make a lot of difference to the contours whereas
leaving out M31 severely limits the statement that can be made about
the mass of the system. In the $\Omega_0 = 0.99$ universe, for
example, we can only say that the combined mass of the system is
probably less than $9 \times 10^{12} M_{\sun}$.

Figure 3 shows the same contours as figure 2 but for the case of an
older universe. The masses generally decrease in this case, because
less mass is needed to produce the same results when gravitation works
over a longer period in time. 
{\vskip 5pt}
\placefigure{fig-3}
\input epsf
\def\figureps[#1,#2]#3.{\bgroup\vbox{\epsfxsize=#2
    \hbox to \hsize{\hfil\epsfbox{#1}\hfil}}\vskip12pt
    \small\noindent Figure#3. \def\par{\endgraf\egroup\vskip12pt}}

\figureps[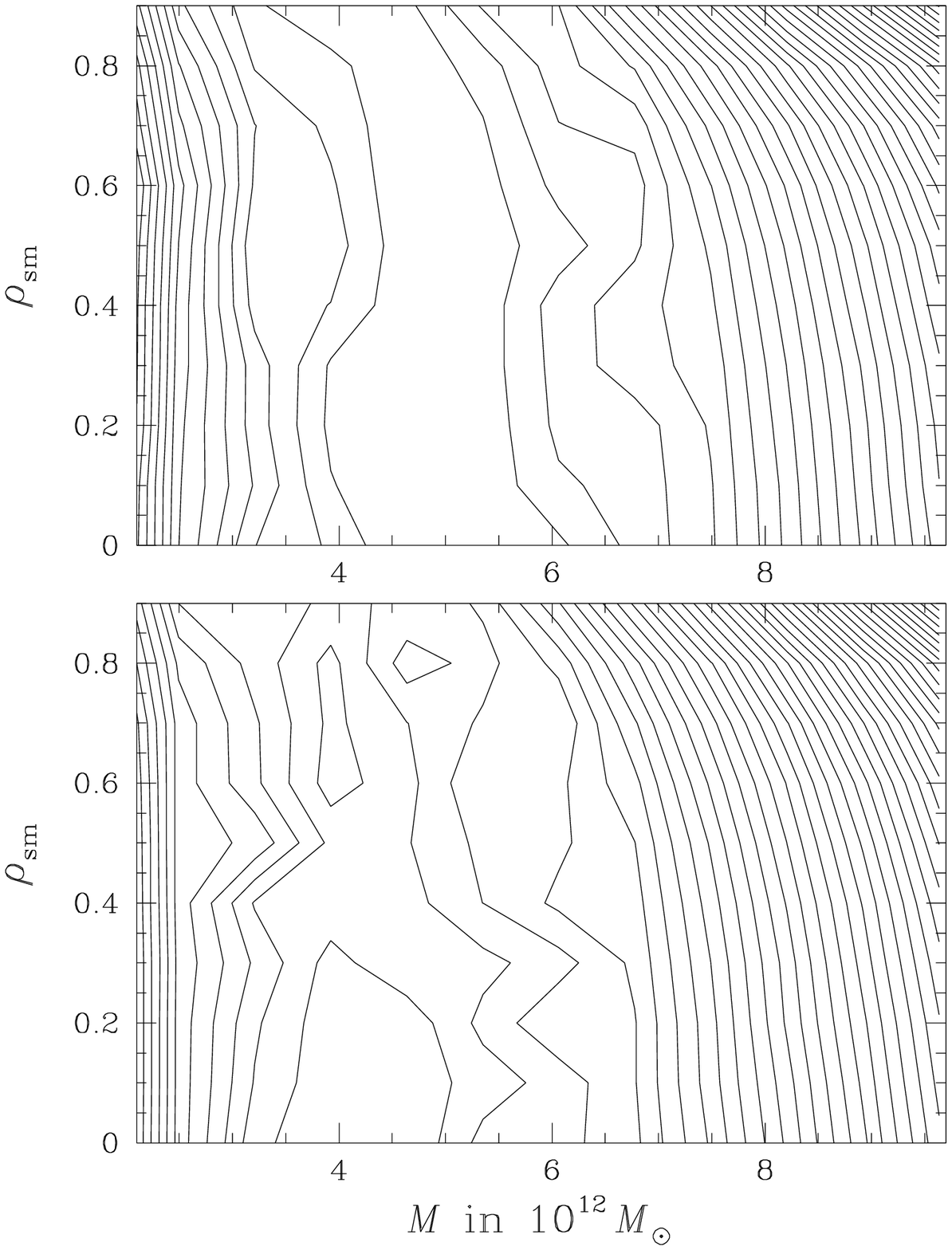,.9\hsize] 3. As for figure 2 but with
$\kappa = 1.5$.

Both figure 2 and 3 were produced from a set of solutions using 8
basis functions. We experimented with different numbers of basis
functions but the results remain the same.

Our results indicate that most of the physics are included in our
model if not all of them. Hence we opted for a likelihood analysis.
The analysis shows the system dominated by M31, which makes our model
not so very different from the one used for the timing argument.
However, if we leave M31 out of the analysis, the resulting contours
are at least consistent with figures 2 and 3. 
There seems to be some slight preference for mass to cluster with the
galaxies, but the contours are not really conclusive.

\section{Discussion}

The $N$-body check proves that the code works and is ready to be
extended to larger systems. The main problem with a system like the
Local Group is 
the occurrence of multiple solutions. We have no guarantee that the
solutions we find are the real ones, even if they fit all redshifts
perfectly and predict distances which are not too far out from the
measured ones. In fact, some tests suggests that at least for certain
masses, several possible solutions are very close to each other so
it is easy to pick the wrong one. This problem will not occur in larger
systems, so they should in fact be easier to deal with.

In future work, two issues will need to be addressed:
First, we need an approximate and more efficient way of
doing the force calculations in equations \ref{reals} and
\ref{reds}. The direct sum that we have used needs to be replaced by a
standard $N$-body method.
Second, we need more efficient convergence. As mentioned above, the
main difficulty lies in preventing the coefficients from finding
two-cycles instead of fixed points of the iteration.
A faster yet still robust method for dealing with this problem is needed.

Even when these problems are solved, the variational calculations will
never have as many particles as $N$-body simulations. The reason for
pursuing this method is that unlike the $N$-body calculation, which
can only reproduce the current state of the system in a statistical
sense, the variational method can fit the current state exactly.

{\vskip 10pt}
\acknowledgments
The authors would like to thank Alan Whiting and James Binney for many
helpful comments and suggestions. I. S. acknowledges a PPARC studentship and an
Oriel College graduate scholarship.

\end{document}